\documentclass[manuscript,screen,nonacm]{acmart}

\AtBeginDocument{%
  }

\usepackage[T1]{fontenc}

\usepackage{amsmath,mathtools}
\usepackage{booktabs}
\usepackage{graphicx}
\usepackage{multirow}
\usepackage{array}
\usepackage{xspace}
\usepackage{tikz}
\usetikzlibrary{positioning}
\usepackage[ruled,vlined,boxed,linesnumbered]{algorithm2e}
\usepackage[nameinlink,noabbrev]{cleveref}

\newcommand{\dfg}{\ensuremath{\mathsf{DFG}}}
\newcommand{\efg}{\ensuremath{\mathsf{EFG}}}

\newcommand{\act}{\ensuremath{\mathcal{A}}}

\newcommand{\covers}{\ensuremath{\prec^{-}}}
\newcommand{\incomp}{\ensuremath{\parallel}}

\begin{document}

\title{Inductive Process Discovery from Partially Ordered Event Data}

\author{Humam Kourani}
\email{humam.kourani@fit.fraunhofer.de}
\orcid{0000-0003-2375-2152}
\correspondingauthor
\affiliation{%
  \institution{Fraunhofer Institute for Applied Information Technology FIT}
  \city{Sankt Augustin}
  \country{Germany}
}

\author{Tom Breuer}
\email{tom.breuer@rwth-aachen.de}
\orcid{0009-0005-6807-6604}
\affiliation{%
  \institution{RWTH Aachen University}
  \city{Aachen}
  \country{Germany}
}

\author{Gyunam Park}
\email{gyunam.park@fit.fraunhofer.de}
\orcid{0000-0001-9394-6513}
\affiliation{%
  \institution{Fraunhofer Institute for Applied Information Technology FIT}
  \city{Sankt Augustin}
  \country{Germany}
}

\author{Wil M. P. van der Aalst}
\email{wvdaalst@pads.rwth-aachen.de}
\orcid{0000-0002-0955-6940}
\affiliation{%
  \institution{RWTH Aachen University}
  \city{Aachen}
  \country{Germany}
}


\begin{abstract}
  The Inductive Miner (IM) family is a prominent class of process discovery techniques, combining efficient recursive decomposition with soundness-by-construction guarantees. However, IM techniques usually assume traces to be totally ordered sequences of activity occurrences. This assumption is convenient, but can introduce systematic bias: activities may have durations, events may share coarse timestamps, or the data may constrain only some event pairs. Forcing such executions into arbitrary sequences hides inherent concurrency and may introduce sequential dependencies that were never observed as causal constraints. Partial orders provide a more faithful representation, but integrating them into IM discovery is challenging because standard abstractions are sequence-based; directly reusing them would require linearizing each partial order, which becomes prohibitively expensive under high concurrency. We introduce a lifting of IM discovery from total orders to partially ordered traces. Instead of redesigning the miner and its cut detection logic, we redefine the trace abstraction layer and the recursive projections to operate directly on partial orders. The approach is conservative over totally ordered traces, avoids linearization explosion, and preserves the recursive structure and guarantees that make IM attractive. Experimental results show that the proposed lifting avoids the combinatorial overhead of linearization, reduces sensitivity to arbitrary tie-breaking in timestamped event data, and allows process behavior to be learned from fewer observations by preserving concurrency at the trace level.
\end{abstract}




\maketitle

\section{Introduction}\label{sec:introduction}

Process discovery is one of the central tasks of process mining. Given event data recorded by information systems, the goal is to derive a process model that describes the observed behavior and can subsequently be used for communication, conformance checking, simulation, automation, and improvement initiatives \cite{DBLP:conf/bpm/AalstAM11}. Among the many discovery paradigms proposed over the years, the Inductive Miner has become a reference point because it combines scalability with formal guarantees. It recursively decomposes the event log using cuts over directly-follows abstractions and constructs a hierarchical model that is sound by construction \cite{DBLP:series/lnbip/Leemans22}. This idea has also been lifted to more expressive intermediate notations, such as the Partially Ordered Workflow Language (POWL), where partial orders and choice graphs can capture non-block-structured concurrency and decisions while retaining the inductive discovery scheme \cite{DBLP:conf/bpm/KouraniZ23,DBLP:journals/is/KouraniZSA25,DBLP:conf/bpm/KouraniPA25}.

Despite these advances, the input side of most inductive discovery algorithms remains predominantly sequential. An event log is typically represented as a multiset of traces, and each trace is a sequence of activity labels. This representation is simple and efficient, but it silently assumes that all relevant events in a case can be placed in one reliable total order. This assumption is often questionable. Activities may overlap in time. Start and complete lifecycle transitions may show that one task was still running while another was started \cite{DBLP:conf/bpm/LeemansFA15}. Several events may share the same timestamp because of coarse temporal granularity. During event abstraction, analysts may intentionally discard accidental ordering evidence and keep only robust precedence constraints. In all these situations, a partially ordered trace is the more faithful representation: some event pairs are ordered, while other pairs are intentionally left incomparable.

The practical impact of forced total ordering is not merely notational. If two activities are concurrent but the log stores one arbitrary sequence per case, discovery sees an artificial order. When the arbitrary order is consistent, the discovered model may impose a false sequential dependency. This problem has motivated a growing body of work on partially ordered event data, partial-order variants, and partial-order-based process mining \cite{DBLP:journals/kais/LeemansZL23}. However, there is still a gap between the availability of partial-order event data and the reuse of mature, scalable inductive discovery machinery. 

A straightforward approach is to enumerate all linearizations of each partially ordered trace and feed the resulting sequential log into an existing miner. This solution is conceptually attractive because it requires no change to the miner, but it comes with substantial computational overhead. A partial order with a concurrency of size $k$ already represents $k!$ permutations, so the number of generated traces can grow factorially with the degree of concurrency. Enumerating these permutations can also change the statistical meaning of the log: one observed execution with many concurrent events may dominate the frequency counts merely because it has many linear extensions. Sampling linearizations reduces the overhead, but introduces variance and may miss orderings that are important for detecting concurrency.

\begin{figure}[!t]
\includegraphics[width=0.8\textwidth]{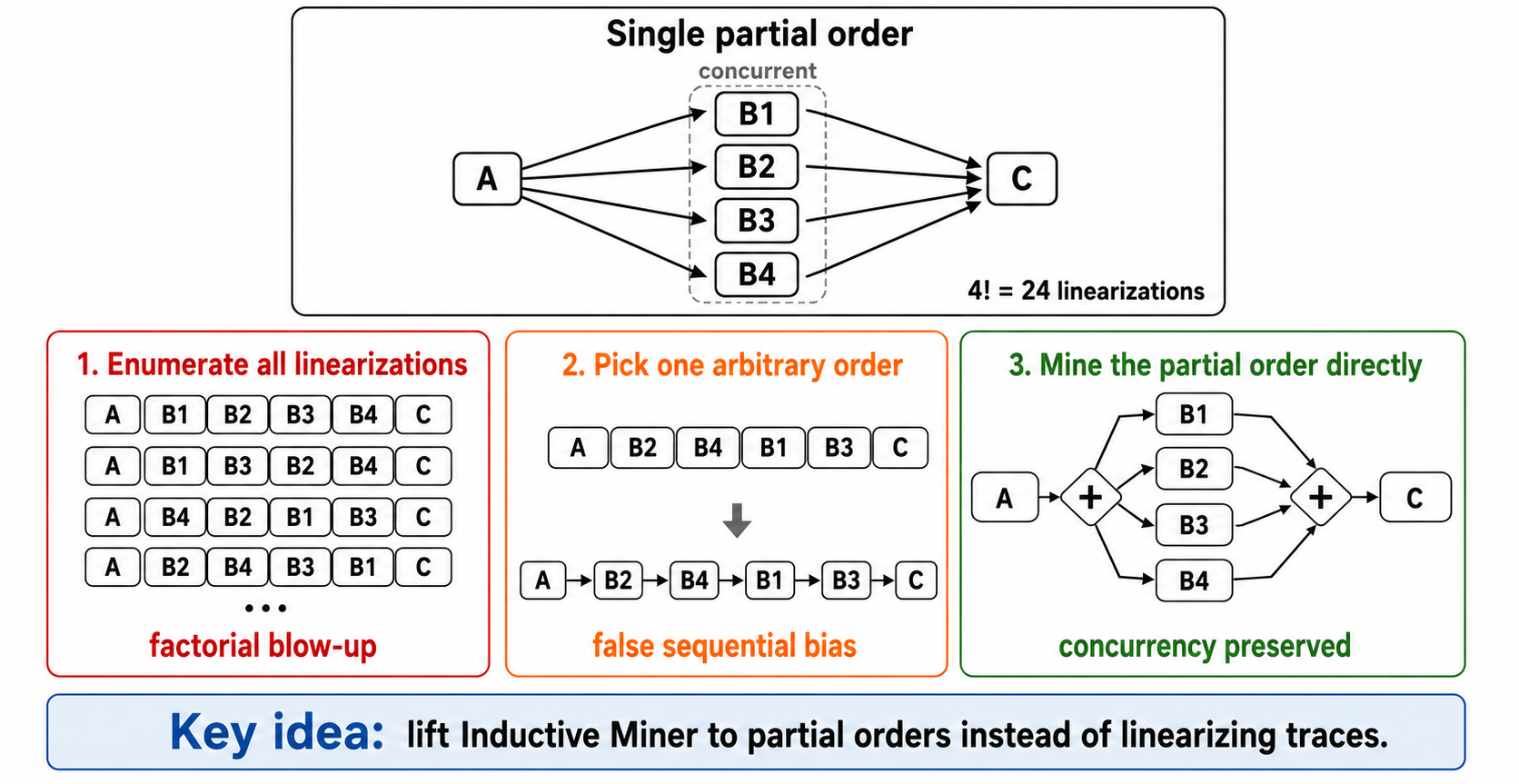}
\caption{Overview of discovery from partially ordered traces. The top panel shows a single partially ordered trace in which activity $A$ is followed by four concurrent activities $B_1,B_2,B_3,B_4$, which are all followed by $C$. This single partial order compactly represents $4!=24$ possible linearizations. Enumerating all linearizations leads to factorial growth, while selecting one arbitrary linearization may introduce false sequential bias. Our approach lifts Inductive Miner to operate directly on partially ordered traces, preserving concurrency in the discovered model.}
\label{fig:pot-overview}
\end{figure}

\Cref{fig:pot-overview} illustrates the core challenge and the proposed solution: instead of linearizing partially ordered traces, we lift the abstraction layer of the Inductive Miner to partial orders. The central design choice is simple: cuts should still be detected over directly-follows and eventually-follows style abstractions, but these abstractions should be computed from partial orders rather than sequences. Since IM is recursive, we also lift projection: after a cut is detected, each sublog is obtained by taking the induced sub-order on the events of the corresponding activity block. Thus, partial-order information is preserved throughout the recursion, not only in the initial abstraction.

We provide an implementation of the proposed approach in the Python library POWL (\url{https://pypi.org/project/powl/}) and evaluate it from three complementary perspectives. The experiments show that direct discovery from partially ordered traces preserves concurrency evidence without the combinatorial overhead of
enumerating linearizations, reduces sensitivity to arbitrary ordering choices in timestamped event data, and can capture the behavior of a concurrent process from fewer observed executions than discovery from sequential traces.

The remainder of the paper is structured as follows. \Cref{sec:related-work} reviews related work, while \Cref{sec:preliminaries} introduces the required concepts and notation. \Cref{sec:method} presents the proposed discovery approach, including the lifted abstraction and projection mechanisms. \Cref{sec:evaluation} evaluates the approach. Finally, \Cref{sec:conclusion} summarizes the findings and outlines directions for future work.

\section{Related Work}\label{sec:related-work}

\paragraph{Inductive process discovery.}
The Inductive Miner discovers block-structured models by recursively detecting cuts over log abstractions \cite{DBLP:series/lnbip/Leemans22}. Its appeal lies in soundness by construction, scalability, and a clear decomposition principle. Several extensions improve robustness, frequency handling, and expressiveness. POWL discovery extends the intermediate representation from process trees to partial orders \cite{DBLP:conf/icpm/KouraniSA23,DBLP:journals/is/KouraniZSA25}, and choice graphs further relax block-structured decision constraints \cite{DBLP:journals/procsci/KouraniPA26}. 

\paragraph{Concurrency and partial-order event data.}
The treatment of inherent concurrency in event data is a growing topic in
process mining \cite{DBLP:journals/kais/LeemansZL23}. Some techniques exploit
lifecycle or interval information to distinguish concurrency from sequential
interleavings, including lifecycle-aware Inductive Miner variants
 \cite{DBLP:conf/bpm/LeemansFA15} and extensions of Split Miner for overlapping
activity intervals \cite{DBLP:conf/icpm/AugustoDR20}. In \cite{DBLP:conf/icpm/RennertPA24}, partial orders are used to evaluate student behavior, where course-taking histories are modeled without imposing unnecessary sequential dependencies. Other work uses richer
behavioral representations such as prime event structures
 \cite{DBLP:conf/apn/DumasG15,DBLP:conf/icpm/Bergenthum19}. Our proposed approach is
orthogonal to these approaches: it assumes that concurrency has already been
represented as partially ordered traces and focuses on consuming such traces
within the IM framework.

\paragraph{Discovery from partial orders.}
Several approaches discover models directly from partially ordered behavior,
including aggregation of instance graphs or causal runs
 \cite{van2005multi,DBLP:journals/topnoc/DongenDA12}, ILP-based discovery
 \cite{DBLP:conf/apn/FolzWeinsteinBDK23}, replay- and synthesis-based mining
 \cite{DBLP:conf/caise/FolzWeinsteinRMBA25}, region-based synthesis \cite{kovavr2024exploratory},
unfolding-based synthesis using independence relations
 \cite{DBLP:conf/atva/LeonRCHH15}, and folding of partially ordered runs
 \cite{DBLP:conf/apn/Nombre11a}. These methods demonstrate the breadth of
partial-order-based discovery, but they generally follow synthesis or
aggregation paradigms rather than recursive IM-style decomposition. Such methods
can also face scalability challenges on large or complex logs \cite{DBLP:journals/kais/LeemansZL23}. 

\paragraph{Partial-order variants in Cortado.}
Partial-order tools such as Cortado support partially ordered event data for variant representation, visualization, querying, and discovery \cite{DBLP:conf/icpm/SchusterSZA21,DBLP:journals/isci/SchusterZZA24,DBLP:journals/softx/SchusterZA23}. In Cortado, concurrent variants are sequentialized before applying process discovery, whereas our proposed approach lifts the abstraction and projection steps of IM so that discovery operates directly on partially ordered traces.

\paragraph{Inductive Miner - Lifecycle.}
The closest related work within the Inductive Miner family is the lifecycle-aware Inductive Miner (IMLC)~\cite{DBLP:conf/bpm/LeemansFA15}. IMLC assumes lifecycle-enriched event logs in which every activity execution is represented by start and completion events. For example, the sequential trace $\langle a_{\mathsf{start}}, b_{\mathsf{start}}, a_{\mathsf{complete}}, b_{\mathsf{complete}}\rangle$ reveals that $a$ and $b$ overlap, since $b$ starts before $a$ completes. IMLC exploits such overlaps to construct concurrency-enriched abstractions for cut detection, while the underlying event trace remains totally ordered and standard sequence-based IM splitting can be retained. Our approach removes this total-order assumption at the representation level: the input consists directly of partially ordered activity occurrences, so two events can be incomparable without being encoded through a particular ordering of lifecycle markers. Lifecycle intervals are therefore only one possible source of such partial orders; they may also originate from coarse timestamps, domain-specific precedence constraints, or other abstraction mechanisms. Since partially ordered traces have no canonical left-to-right order, we additionally lift recursive projection itself so that partial-order information is preserved throughout discovery.

\paragraph{Aggregation-based discovery with POWL}
Another related line of work is the partial-order aggregation-based discovery approach in \cite{DBLP:journals/corr/abs-2509-15346}. This
approach also uses partially ordered traces as input and produces sound
POWL-based models, but it does not use the Inductive Miner framework; it
aggregates partial-order behavior directly. This approach may produce more
complex models as it allows for duplicating activity labels. The two approaches therefore mainly share
their use of POWL as an intermediate representation and their
soundness-by-construction guarantees, while relying on different discovery
frameworks.

\section{Preliminaries}\label{sec:preliminaries}

\subsection{Sequences and Event Logs}

Let $\act$ be a finite universe of activity labels. A \emph{sequence trace} over $\act$ is a finite sequence $\sigma=\langle a_1,\ldots,a_n\rangle$ with $a_i\in\act$ for all $1 \leq i \leq n$. A traditional event log $L$ can be abstracted as a multiset of sequence traces. We write $L(\sigma)$ for the multiplicity of sequence $\sigma \in L$.

Given a sequence $\sigma=\langle a_1,\ldots,a_n\rangle$, the start and end activities are $a_1$ and $a_n$ if $n>0$. The directly-follows support of $\sigma$ is
\[
\dfg(\sigma)=\{(a_i,a_{i+1})\mid 1\leq i<n\}.
\]
The eventually-follows support is
\[
\efg(\sigma)=\{(a_i,a_j)\mid 1\leq i<j\leq n\}.
\]
In practical discovery algorithms these supports are lifted to frequency counters over a multiset of traces. The Inductive Miner uses such abstractions to detect cuts corresponding to control-flow structures \cite{DBLP:series/lnbip/Leemans22,DBLP:conf/bpm/KouraniPA25}.

\subsection{Partially Ordered Traces}

The input considered in this paper generalizes sequence traces. Instead of assuming that all events of a case are totally ordered, we allow partial orders.

\begin{definition}[Partially ordered trace (POT)]\label{def:pot}
A \emph{partially ordered trace} over $\act$ is a tuple
\[
\pi=(E,\prec,\lambda),
\]
where $E$ is a finite set of events, $\prec\subseteq E\times E$ is an irreflexive and transitive relation (i.e., a partial order), and $\lambda:E\to\act$ maps events to activity labels. If neither $e\prec f$ nor $f\prec e$ holds, then $e$ and $f$ are \emph{incomparable}, written $e\incomp f$. A \emph{partially ordered event log (POT log)} is a multiset of partially ordered traces.
\end{definition}

The relation $\prec$ captures forced precedence constraints. Incomparable events may represent intentionally preserved concurrency or uncertainty. For a POT $\pi=(E,\prec,\lambda)$, the \emph{transitive reduction} (also called \emph{cover relation}) is defined as
\[\covers = \{(e, f) \mid e\prec f \ \wedge \ \nexists g\in E: e\prec g \ \wedge \ g\prec f\}.\]
The minimal and maximal events are
\[
\min(\pi)=\{e\in E\mid \nexists f\in E: f\prec e\},
\quad
\max(\pi)=\{e\in E\mid \nexists f\in E: e\prec f\}.
\]

A \emph{topological ordering} of a finite partial order $(E,\prec)$ is a sequence
$\langle e_1,\ldots,e_n\rangle$ containing each event in $E$ exactly once such that,
for all $e_i,e_j\in E$, if $e_i\prec e_j$, then $i<j$. 

\begin{definition}[Linearization]\label{def:linearization}
Let $\pi=(E,\prec,\lambda)$ be a partially ordered trace. A sequence
$\sigma=\langle \lambda(e_1),\ldots,\lambda(e_n)\rangle$ is a
\emph{linearization} of $\pi$ if $\langle e_1,\ldots,e_n\rangle$ is a
topological ordering of $(E,\prec)$. 
The set of all linearizations of $\pi$
is denoted by $\operatorname{Lin}(\pi)$.
\end{definition}

A partial order compactly represents a possibly large set of total orders. If $k$ events are pairwise incomparable, the partial order has $k!$ linearizations. This is the source of the linearization explosion that our proposed approach avoids.

\subsection{POWL and Inductive Discovery}

POWL is a hierarchical modeling language that combines the advantages of process trees with partial orders \cite{DBLP:conf/bpm/KouraniZ23,DBLP:journals/is/KouraniZSA25}. Recent POWL variants also support choice graphs to represent non-block-structured decision behavior \cite{DBLP:conf/bpm/KouraniPA25,DBLP:journals/procsci/KouraniPA26}. The important aspect for this paper is that POWL discovery already follows the inductive decomposition paradigm: a log abstraction is computed, a cut is detected, the log is projected onto the cut blocks, and the algorithm recursively discovers submodels.

This recursive scheme can be summarized as follows. Given an input object $X$, first test base cases such as an empty log or a single activity. If no base case applies, compute a cut. A cut is a partition or ordered grouping of the activity alphabet together with an operator specification. For example, a sequence cut with groups $A_1,\ldots,A_k$ yields a sequential composition of recursively mined projections. If no cut is found, a fall-through model is returned. The strength of this scheme is that the discovered model is composed recursively and can retain soundness guarantees when the modeling constructs are sound by construction.

The classical version assumes that the input object is a sequence log or a directly-follows graph derived from a sequence log. Our proposal changes the derivation of this input object when traces are partial orders. Note that the abstraction proposed in this paper is not limited to the POWL Inductive Miner: any variant of the inductive miner whose cut detectors consume directly-follows and eventually-follows information can adopt the same lifting principle.

\section{POT-Aware Inductive Mining}\label{sec:method}
This section presents the lifting of inductive discovery from totally ordered traces to partially ordered traces. \Cref{subsec:pot-artifacts} introduces the POT abstraction artifacts used by the miner. \Cref{subsec:pot-projection} then defines the projection operations needed for recursive mining. Finally, \Cref{subsec:algorithmic-summary} summarizes the complete pipeline, including the extraction of partially ordered traces from timestamped event data and the recursive discovery procedure that reuses existing inductive cut detection mechanisms.

\subsection{POT Abstraction Artifacts}\label{subsec:pot-artifacts}
For sequence logs, inductive miners derive their control-flow evidence from log-level artifacts such as start activities, end activities, directly-follows relations, and eventually-follows relations. We lift these artifacts to POT logs by interpreting the partial order of each trace directly, rather than first choosing or enumerating compatible total orders. The resulting artifacts expose the precedence constraints and incomparabilities that are relevant for cut detection.

\subsubsection{Start and end counters} For a sequence trace, there is at most one start and one end activity. For a partially ordered trace there may be several minimal and maximal events. We therefore define start and end evidence as sets, lifted to counters over the log.

\begin{definition}[Start and end counters for POT logs]\label{def:start-end}
Let $L$ be a POT log over $\act$. The \emph{start counter}
$\mathsf{start}_L:\act\to\mathbb{N}$ and the \emph{end counter}
$\mathsf{end}_L:\act\to\mathbb{N}$ are defined as
\[
\mathsf{start}_L(a)
=
\sum_{\pi=(E,\prec,\lambda)\in L}
L(\pi)\cdot
\left|\{e\in\min(\pi)\mid \lambda(e)=a\}\right|,
\]
and
\[
\mathsf{end}_L(a)
=
\sum_{\pi=(E,\prec,\lambda)\in L}
L(\pi)\cdot
\left|\{e\in\max(\pi)\mid \lambda(e)=a\}\right|.
\]
\end{definition}

\subsubsection{Directly-follows evidence}
The directly-follows graph is a central abstraction in Inductive Miner cut detection. For partially ordered traces, we derive this evidence directly from the order relation. Cover relations contribute directly-follows evidence in their forced direction, since they represent immediate precedence constraints in the partial order. Incomparable event pairs contribute evidence in both directions, since either order is compatible with the trace. 

\begin{definition}[Directly-follows graph for POT logs]\label{def:pot-dfg}
Let $L$ be a POT log over $\act$. The \emph{directly-follows graph}
$\dfg_L:\act\times\act\to\mathbb{N}$ is defined as
\[
\dfg_L(a,b)
=
\sum_{\pi\in L} L(\pi)\cdot
\bigl(
\mathsf{cov}_{\pi}(a,b)
+
\mathsf{inc}_{\pi}(a,b)
\bigr),
\]
where, for $\pi=(E,\prec,\lambda)$,
\[
\mathsf{cov}_{\pi}(a,b)
=
\left|
\{(e,f)\in E\times E
\mid e\covers f,\ \lambda(e)=a,\ \lambda(f)=b\}
\right|
\]
and
\[
\mathsf{inc}_{\pi}(a,b)
=
\left|
\left\{
\{e,f\}\subseteq E
\ \middle|\ 
e\incomp f,\ 
\{\lambda(e),\lambda(f)\}=\{a,b\}
\right\}
\right|.
\]
\end{definition}

The rule for incomparable events is the crucial bridge to existing cut detection. In sequence logs, concurrent activities often manifest as both $a\rightarrow b$ and $b\rightarrow a$ across different traces. In a POT log, this evidence is already present within a single trace: the two events are incomparable. Adding both arcs lets existing concurrency cut detectors observe the same bidirectionality without sampling or enumerating permutations.

\subsubsection{Eventually-follows evidence} Some IM variants, especially partial-order cuts, use eventually-follows information to determine whether one block must precede another, whether two blocks are concurrent, or whether they should be merged. We therefore also lift eventually-follows evidence to partial orders.

\begin{definition}[Eventually-follows graph for POT logs]\label{def:pot-efg}
Let $L$ be a POT log over $\act$. The \emph{eventually-follows graph}
$\efg_L:\act\times\act\to\mathbb{N}$ is defined as
\[
\efg_L(a,b)
=
\sum_{\pi\in L} L(\pi)\cdot
\bigl(
\mathsf{ord'}_{\pi}(a,b)
+
\mathsf{inc'}_{\pi}(a,b)
\bigr),
\]
where, for $\pi=(E,\prec,\lambda)$,
\[
\mathsf{ord'}_{\pi}(a,b)
=
\left|
\{(e,f)\in E\times E
\mid e\prec f,\ \lambda(e)=a,\ \lambda(f)=b\}
\right|
\]
and
\[
\mathsf{inc'}_{\pi}(a,b)
=
\left|
\left\{
\{e,f\}\subseteq E
\ \middle|\ 
e\incomp f,\ 
\{\lambda(e),\lambda(f)\}=\{a,b\}
\right\}
\right|.
\]
\end{definition}

Comparable event pairs contribute eventually-follows evidence in their forced direction. Incomparable event pairs contribute evidence in both possible directions. 

\subsubsection{Discussion: Frequency Semantics}
Frequency handling for incomparable events is a design choice, and several reasonable normalizations are possible. For example, one could split the weight of an incomparable pair across the two directions, average over all linearizations, or use another weighting scheme tailored to a specific noise-filtering strategy. In our approach, we use a case-based maximal support semantics: a trace with multiplicity $m$ contributes weight $m$ to each forced relation, and if two events are incomparable, it contributes weight $m$ to both possible directions. This choice reflects the fact that, if the partial order were replaced by one arbitrary compatible total order, either orientation of an incomparable pair could be selected by an admissible tie-breaking rule; hence, under a normalized variant that preserves the original case multiplicity rather than expanding the trace into all linearizations, either direction could receive the full support $m$. Our abstraction records this maximal normalized support for each direction, while avoiding the factorial bias of all-linearization counting. Other choices can certainly be justified, and they are orthogonal to our main contribution: any frequency normalization can be applied without changing the overall architecture.

\subsection{Projection of Partially Ordered Traces}\label{subsec:pot-projection}

Recursive discovery requires not only detecting a cut, but also constructing the
sublogs on which the recursive calls are applied. In the classical Inductive
Miner, projection is operator-dependent: different cut types use different
projection rules because they make different assumptions about how activities
occur in a trace. We follow the same principle for partially ordered traces. We define two projection schemes, corresponding to the two
POWL cut types used by the miner. We use \emph{block projection} for partial-order constructs and
\emph{visit-projection} for choice-graphs. Basic process-tree cuts can be handled through these
POWL constructs: sequence and concurrency cuts can be represented as partial
orders, whereas exclusive-choice and loop cuts can be represented by choice
graphs.

\begin{definition}[Block projection]\label{def:block-projection}
Let $\pi=(E,\prec,\lambda)$ be a partially ordered trace and let
$A\subseteq\act$. The \emph{block projection} of $\pi$ on $A$ is the induced
subtrace
\[
\pi\!\upharpoonright_A^{\mathsf{blk}}
=
(E_A,\prec_A,\lambda_A),
\]
where
\[
E_A=\{e\in E\mid \lambda(e)\in A\},\qquad
\prec_A=\prec\cap(E_A\times E_A),\qquad
\lambda_A=\lambda|_{E_A}.
\]
\end{definition}

Block projection keeps all events of the selected activity group in one
projected trace. It is used for partial-order cuts, where each cut block is
interpreted as one component of the same execution. The projection therefore
preserves the ordering constraints among all events belonging to that component.

For choice-graph cuts, block projection is not sufficient. A choice graph may
contain cycles, and therefore the same activity group may be visited several
times within one trace. If all events with labels in the group were collected
into one induced sub-order, distinct visits could be merged into a single
recursive instance. The projection used for choice graphs must therefore first
identify which selected events belong to the same visit.

We identify such visits by a graph construction over the selected events. The
graph connects selected events that should remain in the same projected trace:
events that are locally adjacent in the partial order, and events that are
incomparable. The former preserves local ordering structure, whereas the latter
keeps concurrent selected events together. Selected events that are only related
through events outside the group are not connected directly.

\begin{definition}[Projection graph]\label{def:projection-graph}
Let $\pi=(E,\prec,\lambda)$ be a partially ordered trace and let
$A\subseteq\act$. The \emph{projection graph} of $\pi$ on $A$ is the undirected
graph
\[
\Gamma_A(\pi)=(E_A,R_A),
\]
where
\[
E_A=\{e\in E\mid \lambda(e)\in A\}
\]
and
\[
\{e,f\}\in R_A
\quad\text{iff}\quad
e\neq f
\text{ and }
\bigl(e\covers f \ \vee\ f\covers e \ \vee\ e\incomp f\bigr).
\]
The connected components of $\Gamma_A(\pi)$ are denoted by
$\mathsf{Comp}_A(\pi)$.
\end{definition}

Each connected component of the projection graph yields one projected partially
ordered trace.

\begin{definition}[Visit projection]\label{def:visit-projection}
Let $\pi=(E,\prec,\lambda)$ be a partially ordered trace and let
$A\subseteq\act$. The \emph{visit projection} of $\pi$ on $A$ is the multiset
\[
\pi\!\upharpoonright_A^{\mathsf{vis}}
=
\bigl[
(C,\prec\cap(C\times C),\lambda|_C)
\mid
C\in\mathsf{Comp}_A(\pi)
\bigr].
\]
\end{definition}

Visit projection is used mainly for choice-graph cuts. Our implementation also uses the same component-based idea for $\tau$-loop
fall-throughs. These projections differ only in how the projection graph is
constructed before taking connected components. In the ordinary $\tau$-loop
fall-through, connections through start activities are suppressed. In the strict
$\tau$-loop fall-through, connections from end activities back to start
activities are suppressed. Thus, the fall-through cases reuse the same
graph-and-components projection principle, but with edge relations adapted to
the corresponding fall-through semantics.

\subsection{Algorithmic Summary}\label{subsec:algorithmic-summary}

\Cref{alg:pot-extraction,alg:pot-im} summarize the complete pipeline. The preprocessing step constructs a POT log, while the discovery step recursively applies the abstraction and projection machinery introduced in \Cref{subsec:pot-artifacts,subsec:pot-projection}. The construction of the POT log is independent of the miner: users may preprocess event data in any suitable way before discovery. For example, timestamps may be abstracted to the day level, domain-specific ordering constraints may be added, or accidental ordering evidence may be deliberately removed. If lifecycle information is available, e.g., events with start and completion states, then interval-based techniques can be used to derive the partial order. In such settings, different matching strategies can be applied to pair start and completion events. These extraction techniques are orthogonal to the discovery algorithm and can be integrated into the same pipeline as long as they produce partially ordered traces.

In \Cref{alg:pot-extraction}, we use a deliberately simple timestamp-based extraction. For each unique case identifier $c$, every row $(c,a,t)$ gives rise to a fresh event $e$ with label $\lambda_c(e)=a$ and timestamp $\tau_c(e)=t$. The partial order is then derived purely from timestamps: an event $e$ precedes an event $f$ iff $\tau_c(e)+\Delta < \tau_c(f)$, where $\Delta$ is an optional tolerance window. Hence, events with equal timestamps are incomparable when $\Delta=0$, and events whose timestamps are within the tolerance window are also treated as incomparable.

\begin{algorithm}[t]
\DontPrintSemicolon
\KwIn{Event table $T$ with case column, activity column, timestamp column, and tolerance window $\Delta\geq 0$}
\KwOut{POT log $L$}
$L\gets [\ ]$\;
\ForEach{unique case identifier $c$ in the case column of $T$}{
  $E_c\gets\emptyset$, $\prec_c\gets\emptyset$\;
  \ForEach{row $(c,a,t)$ of case $c$}{
    create a fresh event $e$\;
    add $e$ to $E_c$\;
    set $\lambda_c(e)=a$\;
    set $\tau_c(e)=t$\;
  }
  \ForEach{pair of distinct events $e,f\in E_c$}{
    \If{$\tau_c(e)+\Delta < \tau_c(f)$}{
      add $(e,f)$ to $\prec_c$\;
    }
  }
  $\pi_c\gets(E_c,\prec_c,\lambda_c)$\;
  add $\pi_c$ to $L$\;
}
\Return{$L$}\;
\caption{Timestamp-based extraction of a partially ordered trace log.}
\label{alg:pot-extraction}
\end{algorithm}

After extraction, discovery proceeds recursively. \Cref{alg:pot-im} computes the POT-aware start, end, directly-follows, and eventually-follows abstractions defined in 
\Cref{subsec:pot-artifacts}. The discovery variant $v$ specifies an ordered list of decomposition functions. These functions include the standard cut detectors as well as fall-through functions. Each function is applied to the POT abstractions and may return a cut over an activity partitioning. The cut is then translated into the corresponding POWL construct: sequence and concurrency cuts are represented as partial-order constructs, whereas exclusive-choice and loop cuts are represented as choice-graph constructs. This translation determines the projection scheme used in the recursive step. Partial-order constructs use block projection, while choice-graph constructs use visit projection. The procedure is guaranteed to terminate because the configured list of decomposition functions includes a fall-through function that returns a model whenever no ordinary cut is detected. The resulting POT sublogs are then mined recursively.

\begin{algorithm}[t]
\DontPrintSemicolon
\KwIn{POT log $L$; discovery variant $v$}
\KwOut{POWL model $M = \textsc{POT-IM}(L,v)$}

\If{a base case applies}{
  \Return{$\mathsf{baseModel}(L)$}\;
}

$\alpha \gets (\mathsf{start}_L$, $\mathsf{end}_L$, $\dfg_L$, $\efg_L)$\;

\ForEach{cut detection mechanism $f$ configured by $v$}{
  
  \If{$f(\alpha)$ detects a cut $(c; A_1,\ldots,A_n)$}{
    $p \gets$ translate $c$ into an equivalent POWL construct\;

    \eIf{$p$ is a partial-order}{
      $(L_1,\ldots,L_n)\gets
      \mathsf{blockProject}(L;A_1,\ldots,A_n)$\;
    }{
      $(L_1,\ldots,L_n)\gets
      \mathsf{visitProject}(L;A_1,\ldots,A_n)$\;
    }

    \ForEach{$i\in\{1,\ldots,n\}$}{
      $M_i\gets\textsc{POT-IM}(L_i,v)$\;
    }

    \Return{$\mathsf{compose}(p;M_1,\ldots,M_n)$}\;
  }
}
\caption{POT inductive discovery.}
\label{alg:pot-im}
\end{algorithm}


\section{Experimental Evaluation}\label{sec:evaluation}
The evaluation focuses on three central aspects of the proposed lifting of
inductive discovery to partially ordered traces. First, we evaluate scalability
against linearization: our approach should avoid the factorial blow-up caused
by materializing all total orders represented by a partially ordered trace.
Second, we evaluate quality and representation bias on a real-life event log:
sequence-based discovery must commit to an arbitrary tie-breaking order when
events share the same timestamp, whereas our approach can preserve such ties
as incomparabilities. Third, we evaluate the informational value of the
partial-order representation itself in a controlled setting, by investigating
whether preserving concurrency explicitly allows the intended process behavior
to be captured from fewer observed executions than with sequential traces.

We use the default POWL inductive discovery configuration provided
by the POWL package.


\subsection{Scalability on Synthetic Partially Ordered Traces}
\label{subsec:eval-synthetic}
\subsubsection{Setup}
The scalability experiment uses the simple partially ordered pattern from
\Cref{fig:pot-overview}. Each case consists of an initial activity $A$, followed
by a concurrent block of width $k$, followed by a final activity $C$:
\[
A \ ; \ (B_1 \parallel \cdots \parallel B_k) \ ; \ C .
\]

For every value of $k$, the POT log contains one case with one partially ordered
trace. This single trace has $k+2$ events and compactly represents $k!$
sequential executions. We consider $k\in\{2,4,6,8,10\}$, ranging from two
compatible sequential traces for $k=2$ to $10!=3{,}628{,}800$ for $k=10$.

We compare three settings:
\begin{enumerate}\item 
    \textbf{POT-IM}: the proposed method, which consumes the partially ordered trace directly.
    \item \textbf{single linearization}: the sequence-based miner that treats the order of events in the input trace as the actual sequential order.
    \item \textbf{all linearizations}: the sequence-based baseline preceded by a preprocessing step that materializes all topological orderings of the partially ordered trace.

\end{enumerate}

For the all-linearizations baseline, we report the time needed to materialize
the linearized log separately from the subsequent IM discovery time. For all three settings, we
also report whether the concurrent block $B_1,\ldots,B_k$ is detected in the
discovered model.

\subsubsection{Results}
\Cref{tab:synthetic-scalability} shows the scalability results. For the
POT-based approach and the single-linearization baseline, discovery remains
fast across all considered values of $k$. The all-linearizations baseline, 
in contrast, becomes increasingly expensive
as the degree of concurrency grows. At $k=10$, materializing all linearizations takes about
$32.2$ minutes, and the subsequent IM discovery step takes another $10$
minutes.

\begin{table}[t]
\centering
\caption{Scalability on the synthetic concurrency benchmark.}
\label{tab:synthetic-scalability}
\begingroup
\setlength{\tabcolsep}{4pt}
\resizebox{0.8\textwidth}{!}{%
\begin{tabular}{@{}clrrrrc@{}}
\toprule
\multirow{2}{*}{$k$}
& \multirow{2}{*}{Variant}
& \multicolumn{2}{c}{Input log}
& \multicolumn{2}{c}{Runtime (s)}
& \multirow{2}{*}{\begin{tabular}[c]{@{}c@{}}Concurrency detected?\end{tabular}} \\
\cmidrule(lr){3-4}
\cmidrule(lr){5-6}
& & cases & events
& \begin{tabular}[c]{@{}c@{}}linearization\end{tabular}
& \begin{tabular}[c]{@{}c@{}}IM discovery\end{tabular}
& \\
\midrule
\multirow{3}{*}{2}
& POT-IM                     & 1 & 4        & --      & 0.004 & yes \\
& single linearization                & 1 & 4        & --      & 0.007 & no  \\
& all linearizations         & 2 & 8        & 0.006   & 0.004 & yes \\
\midrule
\multirow{3}{*}{4}
& POT-IM                     & 1 & 6        & --      & 0.003 & yes \\
& single linearization                & 1 & 6        & --      & 0.004 & no  \\
& all linearizations         & 24 & 144     & 0.006   & 0.008 & yes \\
\midrule
\multirow{3}{*}{6}
& POT-IM                     & 1 & 8        & --      & 0.008 & yes \\
& single linearization                & 1 & 8        & --      & 0.006 & no  \\
& all linearizations         & 720 & 5{,}760 & 0.164  & 0.060 & yes \\
\midrule
\multirow{3}{*}{8}
& POT-IM                     & 1 & 10       & --      & 0.004 & yes \\
& single linearization                & 1 & 10       & --      & 0.005 & no  \\
& all linearizations         & 40{,}320 & 403{,}200 & 3.802 & 2.351 & yes \\
\midrule
\multirow{3}{*}{10}
& POT-IM                     & 1 & 12       & --      & 0.008 & yes \\
& single linearization                & 1 & 12       & --      & 0.007 & no  \\
& all linearizations         & 3{,}628{,}800 & 43{,}545{,}600 & 1{,}929.244 & 600.539 & yes \\
\bottomrule
\end{tabular}%
}
\vspace{1mm}
\endgroup
\end{table}

Overall, the experiment illustrates how enumerating all
linearizations exposes the same concurrency evidence to the sequence-based miner,
but only at factorial cost. Using only one raw-order sequence is efficient, but
introduces a sequential dependency that is not present in the partially ordered
trace. POT-IM combines the advantages of both alternatives: it preserves the
concurrency information of the partially ordered trace while avoiding the
factorial overhead of materializing all compatible linearizations.

\subsection{Quality and Tie-Breaking Bias on BPI Challenge 2012}
\label{subsec:eval-bpi2012}

\subsubsection{Setup}
The second experiment evaluates the effect of arbitrary timestamp tie-breaking on
a real-life event log. We use the BPI Challenge 2012 event log \cite{bpic12},
because it contains events that share identical timestamps. This experiment is 
not intended to evaluate lifecycle-based partial-order
extraction. Although lifecycle information can in principle be used to derive
interval-based concurrency, most activities in this log are available only with
completion timestamps. We therefore keep completion
events only and construct a POT log using the timestamp-based extraction of
\Cref{alg:pot-extraction} with tolerance window $\Delta=0$. Events of the same
case with identical timestamps are thus treated as incomparable.

We compare this POT representation with two sequential versions of the same log.
Both sequential logs order events by timestamp. Ties are resolved either by
ascending activity label, yielding $L_{\mathsf{asc}}$, or by descending activity
label, yielding $L_{\mathsf{des}}$. Hence, the two sequential logs contain the
same cases, events, activities, and timestamps; they differ only in the
arbitrary order assigned to timestamp ties.

For each input representation, we discover one model without filtering and one
model with filtering threshold $0.8$. Each discovered model is evaluated on both
$L_{\mathsf{asc}}$ and $L_{\mathsf{des}}$ using alignment-based conformance
checking. We use PM4Py \cite{DBLP:journals/simpa/BertiZS23} to compute average trace fitness, the percentage of fitting traces, and
precision. 

\begin{table}[t]
\centering
\caption{Alignment-based conformance quality on BPI Challenge 2012. $\Delta$ denotes the absolute difference between the scores on
$L_{\mathsf{asc}}$ and $L_{\mathsf{des}}$. Lowest $\Delta$ values per filtering
setting and metric are shown in bold.}
\label{tab:bpi2012-results}
\begingroup
\scriptsize
\setlength{\tabcolsep}{4pt}
\resizebox{0.8\textwidth}{!}{%
\begin{tabular}{ll|rrr|rrr|rrr}
\toprule
\multirow{2}{*}{Filtering}
& \multirow{2}{*}{Discovered model}
& \multicolumn{3}{c|}{Average trace fitness}
& \multicolumn{3}{c|}{Fitting traces (\%)}
& \multicolumn{3}{c}{Precision} \\
\cmidrule(lr){3-5}
\cmidrule(lr){6-8}
\cmidrule(lr){9-11}
& & $L_{\mathsf{asc}}$ & $L_{\mathsf{des}}$ & $\Delta$
& $L_{\mathsf{asc}}$ & $L_{\mathsf{des}}$ & $\Delta$
& $L_{\mathsf{asc}}$ & $L_{\mathsf{des}}$ & $\Delta$ \\
\midrule
\multirow{3}{*}{--}
& $\mathrm{Seq\text{-}IM}(L_{\mathsf{asc}})$
& 1 & 0.991 & 0.009
& 100 & 80.6 & 19.4
& 0.220 & 0.253 & 0.033 \\
& $\mathrm{Seq\text{-}IM}(L_{\mathsf{des}})$
& 0.971 & 1 & 0.029
& 64.8 & 100 & 35.2
& 0.252 & 0.219 & 0.033 \\
& $\mathrm{POT\text{-}IM}$
& 1 & 1 & \textbf{0}
& 100 & 100 & \textbf{0}
& 0.174 & 0.178 & \textbf{0.004} \\
\midrule
\multirow{3}{*}{$0.8$}
& $\mathrm{Seq\text{-}IM}(L_{\mathsf{asc}})$
& 0.852 & 0.834 & 0.018
& 26.2 & 26.2 & \textbf{0}
& 0.712 & 0.692 & 0.020 \\
& $\mathrm{Seq\text{-}IM}(L_{\mathsf{des}})$
& 0.812 & 0.842 & 0.030
& 26.2 & 26.2 & \textbf{0}
& 0.696 & 0.651 & 0.046 \\
& $\mathrm{POT\text{-}IM}$
& 0.834 & 0.827 & \textbf{0.007}
& 26.2 & 26.2 & \textbf{0}
& 0.669 & 0.674 & \textbf{0.005} \\
\bottomrule
\end{tabular}%
}
\endgroup
\end{table}

\subsubsection{Results}
\Cref{tab:bpi2012-results} shows that sequence-based discovery is sensitive to
the chosen tie-breaking convention. Without filtering, the model discovered from
$L_{\mathsf{asc}}$ fits $L_{\mathsf{asc}}$ perfectly, but its fitting-trace
score drops to $80.6\%$ on $L_{\mathsf{des}}$. Conversely, the model discovered
from $L_{\mathsf{des}}$ fits $L_{\mathsf{des}}$ perfectly, but drops to
$64.8\%$ on $L_{\mathsf{asc}}$. Thus, the sequential miner may learn ordering
artifacts introduced only by the tie-breaking rule. The POT-based model avoids this dependency. Since timestamp ties are kept incomparable during discovery, the same model obtains perfect average trace
fitness and fitting-trace scores on both sequential realizations. Its precision is lower than that of the
sequence-based models, which is expected: more
permissive concurrency generally reduces precision. However, the POT-based model
is also the most stable with respect to precision, with the smallest precision
difference across the two evaluation logs.

With filtering threshold $0.8$, the absolute scores become more balanced. All
models obtain the same percentage of fitting traces on both sequential logs, and
the model discovered from $L_{\mathsf{asc}}$ achieves the highest absolute
fitness and precision values. Nevertheless, the POT-based model remains the
least sensitive to the tie-breaking convention.  

Overall, the experiment confirms that treating timestamp ties as
incomparabilities reduces the dependence of discovery results on arbitrary
tie-breaking. POT-IM preserves the ordering information supported by the data
while avoiding sequential dependencies introduced only by a chosen
linearization. This comes with an expected loss of precision, because the model
allows behavior corresponding to multiple compatible orderings of tied events.
However, this additional behavior reflects uncertainty already present in the
event data rather than behavior introduced by the discovery algorithm itself.

\subsection{Representation Efficiency under Paired Simulation}
\label{subsec:eval-controlled}

\subsubsection{Motivation}

The real-life tie-breaking experiment in
\Cref{subsec:eval-bpi2012} demonstrates why preserving incomparability can be
useful when the event data themselves do not determine a unique order.
However, such an experiment alone does not isolate whether the advantage comes
from the partial-order representation itself or merely from avoiding a
particular data-quality artifact. We therefore add a controlled experiment in
which both input representations are generated from exactly the same process
executions. The only difference is whether concurrency is retained explicitly
or linearized.

The intuition is straightforward. If two activities are concurrent, one POT
can encode this directly through incomparability. A sequence cannot: it must
show either one activity before the other or vice versa. A sequence-based
miner therefore needs multiple executions with sufficiently diverse
interleavings before bidirectional ordering evidence becomes visible. The
experiment tests whether this difference translates into faster coverage
of the behavior generated by the reference model.

\subsubsection{Reference Model and Paired Simulation}

\begin{figure}[t]
    \centering
    \includegraphics[width=\textwidth]{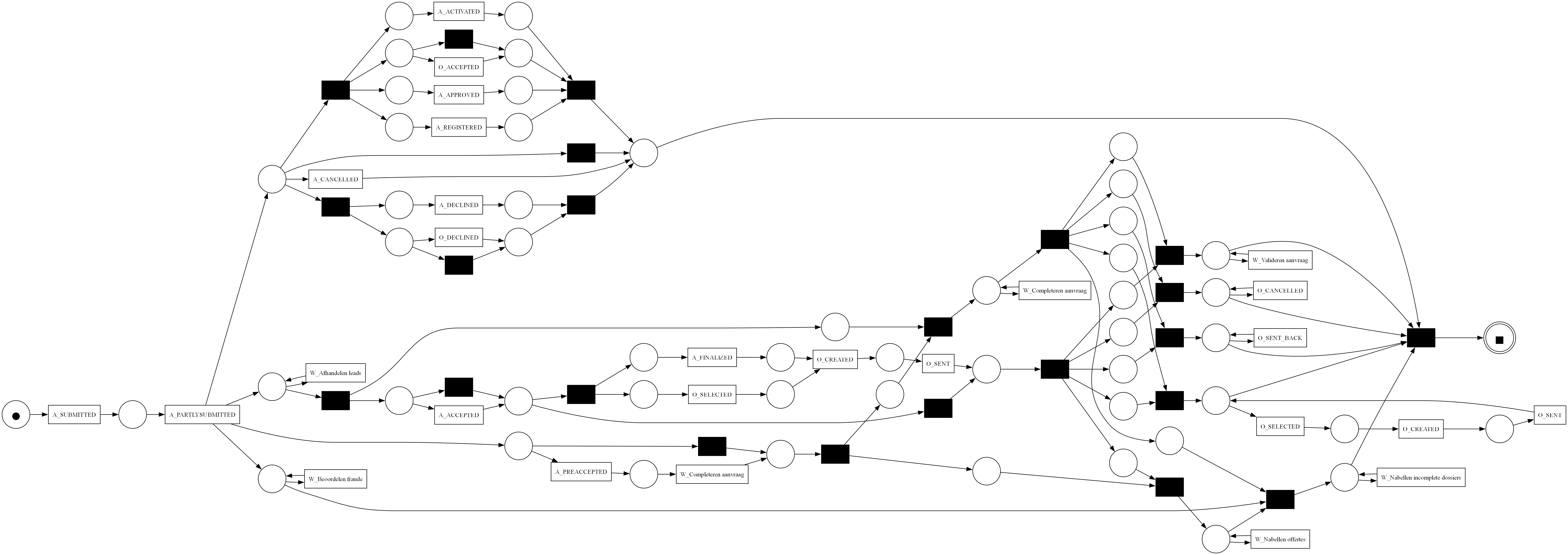}
    \caption{Fixed reference model used for the controlled paired-simulation
    experiment. It was obtained from BPI Challenge 2012 log \cite{bpic12} using the
    partial-order aggregation-based discovery approach
    \cite{DBLP:journals/corr/abs-2509-15346} and is used here only as an
    assumed reference process for simulation.}
    \label{fig:controlled-reference-model}
\end{figure}

We use the BPI Challenge 2012 event log \cite{bpic12} to obtain a
concurrency-rich model with the partial-order aggregation-based
discovery approach \cite{DBLP:journals/corr/abs-2509-15346}. We then
\emph{freeze} the resulting POWL model and use it solely as an assumed
reference process for controlled simulation. The corresponding Petri net is
shown in \Cref{fig:controlled-reference-model}. Importantly, the experiment
does not claim that this model is the true process underlying BPI Challenge
2012; its role is to provide a realistic and concurrency-rich process whose
behavior is fully controlled during simulation.

From this fixed model, we simulate $1{,}000$ non-empty executions. Each sampled execution is first retained as a
partially ordered trace $\pi_i$ and is then randomly topologically sorted to
obtain one compatible sequential trace $\sigma_i$. Consequently,
\[
    \sigma_i \in \operatorname{Lin}(\pi_i)
    \qquad\text{for every } i\in\{1,\ldots,1000\}.
\]
The POT and sequential logs therefore contain the same sampled branch choices,
optional behavior, repetitions, and executed activities. Their only systematic
difference is the representation of concurrent events: the POT keeps them
incomparable, whereas the sequential trace commits to one random compatible
order. We use a fixed random seed for reproducibility.

We then increase the amount of discovery input gradually. For each $n \in \mathcal{N}
=
\{1,2,\ldots,30,40,50,60,70,80,90,100\}$,
we discover one model from the first $n$ partial-order traces and one model
from the corresponding $n$ sequential traces. Thus, at every input size, both
discovery runs are based on equally many executions of the same reference
process.

We report two complementary quality measures. First, every discovered model is
aligned against the same full sequential simulation log containing all
$1{,}000$ traces, and we report the percentage of traces that fit
\emph{perfectly}. Using one common sequential evaluation log avoids giving the
POT-discovered models a special conformance representation and makes this a
strict, easily interpretable behavioral-coverage measure. Second, we compare each
discovered model directly with the fixed reference model using the behavioral
similarity measure implemented in PM4Py \cite{DBLP:journals/simpa/BertiZS23}.
Intuitively, this captures
how closely the behavior of the discovered model matches the behavior of the
reference process, complementing the trace-based perfect-fit measure with a
model-level comparison.

\subsubsection{Results}

\begin{figure}[!t]
    \centering
    \begin{minipage}[t]{0.49\textwidth}
        \centering
        \includegraphics[width=\linewidth]{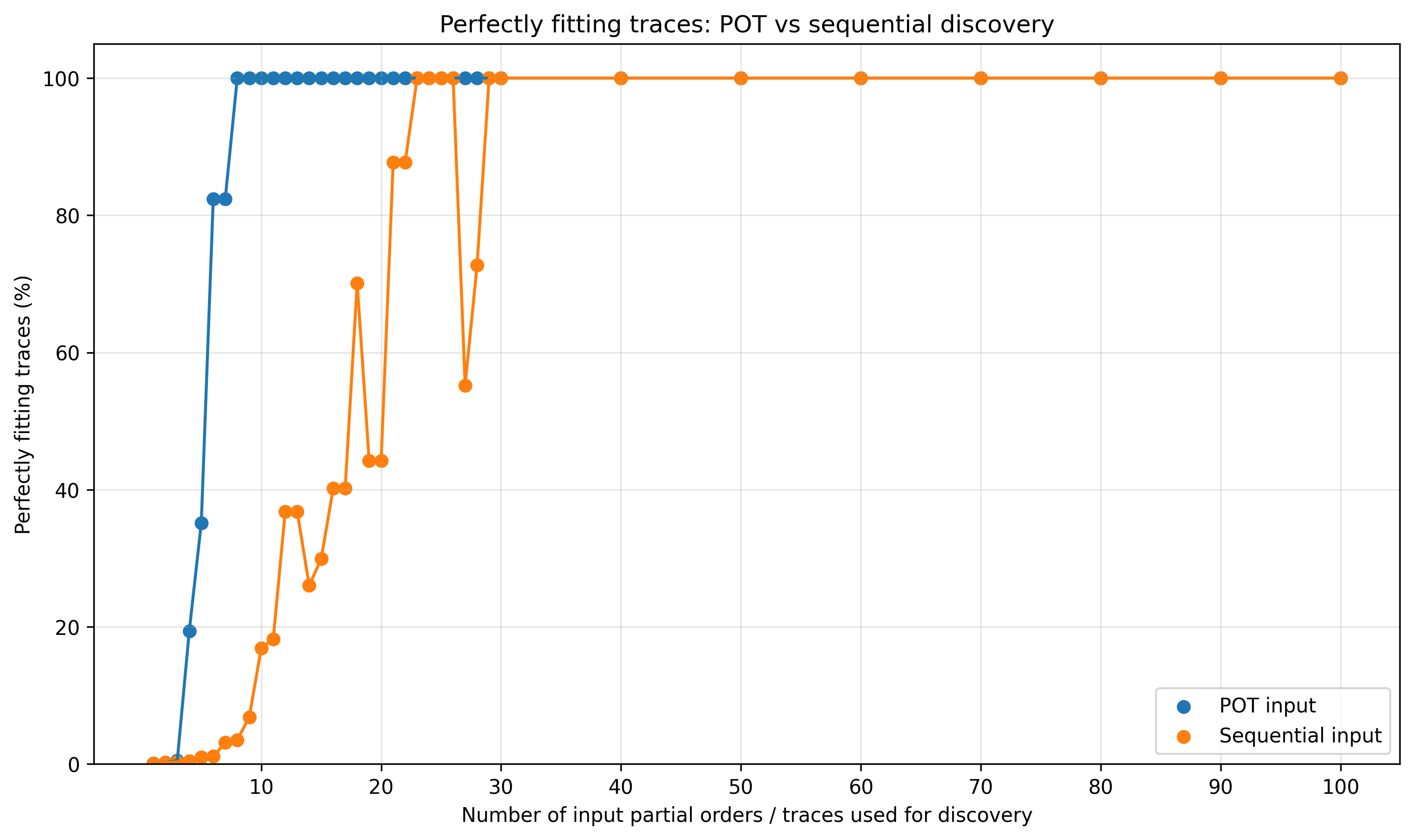}
        \vspace{-1mm}

        {\small\textbf{(a)} Perfectly fitting simulated traces.}
    \end{minipage}\hfill
    \begin{minipage}[t]{0.49\textwidth}
        \centering
        \includegraphics[width=\linewidth]{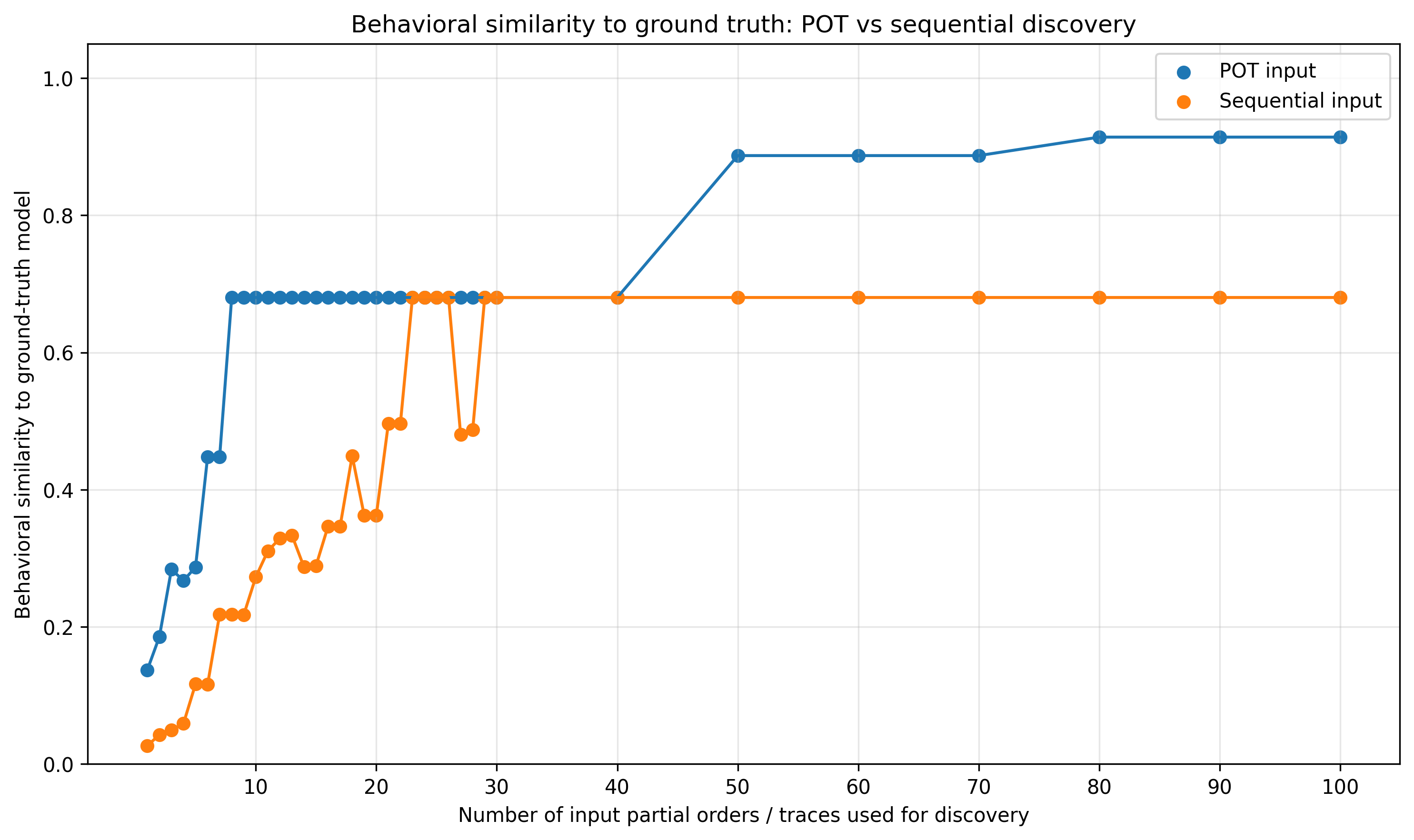}
        \vspace{-1mm}

        {\small\textbf{(b)} Behavioral similarity to the reference model.}
    \end{minipage}
    \caption{Controlled paired-simulation experiment. For every $n$, the POT
    and sequential discovery inputs describe the same $n$ simulated
    executions; only the representation differs. \textbf{(a)} Percentage of
    the common $1{,}000$-trace sequential simulation log that fits the
    discovered model perfectly. \textbf{(b)} Behavioral similarity between
    the discovered model and the fixed reference model. The ``ground truth''
    wording inside panel~(b) refers to this fixed assumed reference model.}
    \label{fig:controlled-quality}
\end{figure}

\Cref{fig:controlled-quality} shows a pronounced sample-efficiency advantage
for the POT representation. The percentage of perfectly fitting traces rises
rapidly for POT-based discovery: with $n=9$ input POTs, all $1{,}000$
sequential reference traces already fit perfectly, and this $100\%$ coverage
is retained for every larger input size considered. Sequential discovery
requires substantially more observations. It first reaches $100\%$ perfectly
fitting traces only around $n=24$ and is not yet stable: the score drops again
for the inputs around $n=27$--$28$ before returning to $100\%$ from $n=29$
onward.

The model-level comparison strengthens this observation. POT-based discovery
reaches a behavioral similarity of approximately $0.68$ after only eight
observations. The sequential variant reaches roughly the same level only after
about two dozen observations. More importantly, the POT-discovered model
continues to improve as additional executions are observed: the similarity is
approximately $0.88$ at $n=50$ and approximately $0.91$ from $n=80$ onward.
In contrast, the sequentially discovered models remain at approximately
$0.68$ for the larger input sizes shown.

The two panels capture complementary aspects of the same effect. Perfectly
fitting traces provide the strongest evidence that the POT representation
allows the miner to cover the behavior generated by the reference process from
few executions. Behavioral similarity then adds a model-level perspective:
even once both approaches can replay all sampled reference traces, the models
discovered from POTs remain behaviorally closer to the process from which the
executions were generated. 

Crucially, this experiment isolates a mechanism that is different from
correcting noisy or ambiguous event data. The sequential traces are not
incorrect: each is a valid linearization of exactly the same underlying
execution represented by its paired POT. Nevertheless, the
sequence-based miner must reconstruct this concurrency indirectly from
variation across several traces, whereas the POT-based miner can observe it
directly within one execution. The earlier replay coverage and higher
model-level similarity therefore provide evidence that partial orders can be
more information-efficient observations for process discovery, not only a
remedy for arbitrary timestamp tie-breaking.

\section{Conclusion}\label{sec:conclusion}

This paper presented a lifting of inductive process discovery from totally ordered traces to partially ordered traces. The proposed approach enables Inductive Mining discovery to operate directly on partial-order event data, avoiding the need to select arbitrary linearizations or enumerate all compatible total orders. To achieve this, we redefined the abstraction layer used by inductive discovery, including start and end activities, directly-follows evidence, and eventually-follows evidence, so that both forced precedence constraints and incomparabilities are reflected in the mined behavior. We also introduced projection mechanisms for partially ordered traces, allowing recursive discovery to preserve partial-order information throughout the mining procedure. The resulting POT-aware inductive miner remains compatible with the recursive decomposition principle of existing inductive discovery techniques while addressing the representational bias and combinatorial overhead caused by linearization.

Future work can extend the proposed framework in several directions. First, alternative frequency semantics for incomparable event pairs should be investigated, including normalized, probabilistic, or noise-aware weighting schemes. Second, the current projection mechanisms can be generalized to additional process constructs, enabling a broader range of inductive discovery variants to consume partially ordered traces directly. Finally, more advanced extraction techniques should be integrated, especially lifecycle- and interval-based methods that derive partial orders from overlapping activity executions.

\section*{Acknowledgments}
This work was supported by the Federal Ministry of Research, Technology and Space (BMFTR), Germany (grant 16IS23065).

\bibliographystyle{splncs04}
\bibliography{references}

\end{document}